\documentstyle[aps,epsf]{revtex}  

\begin{document}        

\baselineskip 14pt
\title{MiniBooNE: the Booster Neutrino Experiment}
\author{Andrew O. Bazarko}
\address{Princeton University, Princeton, NJ 08544\footnote{Presented at
{\it DPF'99}, Los Angeles, January 1999.}}
%
\maketitle              

\begin{abstract}        

The Booster Neutrino Experiment at Fermilab is preparing to search for
$\nu_\mu \rightarrow \nu_e$ oscillations. The experiment is designed
to make a conclusive statement about LSND's 
neutrino oscillation evidence. 
The experimental prospects are outlined in light of the current 
results from LSND and KARMEN. 

\
\end{abstract}   	

\section{Introduction}               

Currently, the 
Liquid Scintillator Neutrino Detector (LSND) at the Los Alamos National 
Laboratory is the only accelerator-based neutrino experiment  
to have evidence for neutrino oscillations \cite{church}.  

The Booster Neutrino Experiment (BooNE) at Fermilab is being prepared
to conclusively test these results. 
The experiment will take place at a new neutrino beamline coming
off of the FNAL 8 GeV proton Booster. 
The first phase of BooNE --- MiniBooNE --- will be a single-detector
experiment.  
MiniBooNE will obtain approximately 1000 events
per year if the LSND signal is due to $\nu_\mu \rightarrow \nu_e$ 
oscillations, and will be capable of establishing the signal with 
greater than 5$\sigma$ significance.  
This new experiment expects to be collecting data by the end of 2001.

KARMEN, the 
Karlsruhe Rutherford Medium Energy Neutrino experiment
at the Rutherford Appleton Laboratory, 
is very similar to LSND, has been running since 1990, and 
will be completed in early 2001.  Therefore we discuss 
the results of both KARMEN and LSND to highlight the 
experimental context that MiniBooNE will encounter.

\section{LSND and KARMEN}
 
LSND presented its first evidence for $\bar\nu_\mu \rightarrow \bar\nu_e$
oscillations in 1995 \cite{lsnd1}.  
In the same year, 
KARMEN completed its first phase of searching for 
the same oscillations, but lacked the sensitivity to confirm 
or refute LSND \cite{karmen1}.  After upgrading its cosmic ray veto 
\cite{upgrade}, 
KARMEN resumed data taking in February 1997 and plans to continue 
running through 2001.  
Results based on KARMEN data collected up to April 1998
were available at the time of {\it DPF'99} \cite{karmen2}.  
More recently KARMEN has updated results based on data collected 
through February 1999 \cite{karmen3}.

LSND and KARMEN both search for $\bar \nu_\mu \rightarrow \bar \nu_e$.
Both experiments are at 800 MeV proton accelerators where 
muon-antineutrinos are produced 
from the decay of muons at rest, through the decay chain:
\begin{eqnarray}
\pi^+ & \rightarrow & \mu^+ \nu_\mu  \nonumber \\
   &  &  \hookrightarrow e^+\, \nu_e \, \bar\nu_\mu \hspace*{1.5cm} 
          {\rm \; endpoint \; 52.8 \; MeV} \nonumber
\end{eqnarray}
Both experiments employ liquid scintillator detectors.
Without being able to directly distinguish $e^+$ from $e^-$, 
the experiments distinguish the appearance of $\bar\nu_e$ 
from the presence of $\nu_e$
by correlating the 
electron-type track in position and time with the photon
from an associated neutron capture reaction.  Neutrons
captured on protons produce 2.2 MeV photons:
\begin{eqnarray}
 \bar\nu_e p & \rightarrow e^+ & n  \nonumber \\
 & & n \, p \rightarrow d \, \gamma {\rm \: (2.2 \; MeV)} \nonumber
\end{eqnarray}
 
There are, of course, differences between the two experiments \cite{yellin}.
The LSND detector is three times the size of the one at KARMEN, 
167 tons versus 56 tons; and LSND's proton exposure of more than 25000 C 
outpaces KARMEN's post-upgrade expectation of 9000 C. 
LSND is positioned 30 m from the neutrino 
source, compared with 17.6 m for KARMEN.  By being 
further from the target, LSND gains sensitivity to lower $\Delta m^2$
at the expense of reduced neutrino flux.
LSND is a single tank 
whereas KARMEN is segmented into 512 modules, and the concentration
of scintillator in LSND is lower than at KARMEN.    
KARMEN therefore has better position and energy resolution, whereas
LSND can measure the track direction and use Cherenkov rings for
particle identification.  Finally, because the target at LSND 
includes a drift space and the detector is positioned downstream, 
LSND is able to also search for the charge conjugate 
$\nu_\mu \rightarrow \nu_e$ oscillation using neutrinos from $\pi^+$
decay in flight \cite{lsnddif}.

\begin{figure}[b,t]
\centerline{\epsfxsize 3.0 truein \epsfbox{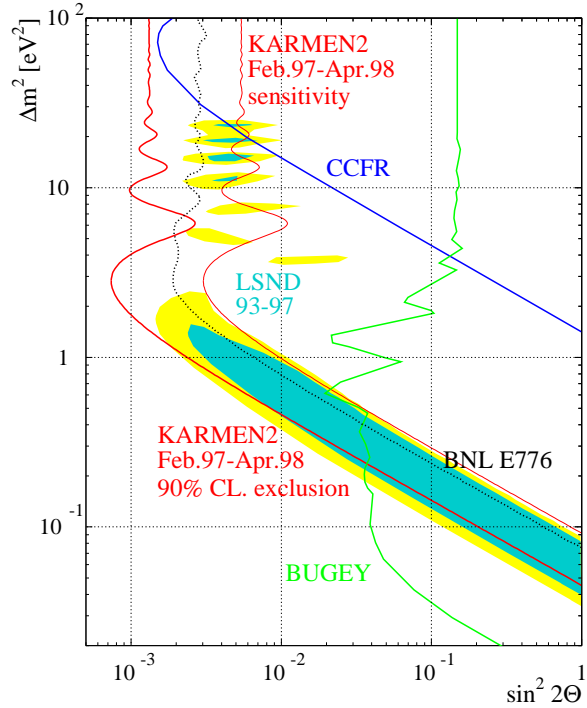}}
\vskip .2 cm
\caption[]{
\label{fig:karmen}
\small KARMEN 90\% confidence level limit and sensitivity based on data 
 collected through April 1998.  The LSND favored region 
 and limits from other experiments are also shown.}
\end{figure}

Based on analysis of data collected  through April 1998, 
which was a proton exposure of 2900 C, 
KARMEN observes 0 candidate events while
expecting $2.88 \pm 0.13$ background events.  With this data, the KARMEN 
sensitivity to an LSND-type signal is on the order of 1 event. 
Figure \ref{fig:karmen} shows the limit 
derived from this experimental observation overlaying the LSND 
favored region.  Also shown is KARMEN's sensitivity, the limit 
if only the background expectation were observed.

Noting that KARMEN expects three times more data, 
a glance at Figure \ref{fig:karmen} raises the question, 
At this rate won't KARMEN soon rule out LSND? 
Well, not so fast.  The difference between KARMEN's sensitivity and limit
curves merits scrutiny.  The limit here benefits from 
the non-observation of even the expected background.  KARMEN will keep 
running, and assuming that events are seen in the future, which 
would be more consistent with the background expectation, KARMEN's 
limit will necessarily move closer to its
sensitivity contour.  The sensitivity will improve 
with more data and with a more sophisticated likelihood 
analysis.  Nevertheless, KARMEN will lack the sensitivity 
to rule out or confirm LSND.

The situation is illustrated by
KARMEN's recently updated results, based on data gathered through
February 1999, about half of its ultimate total.  With a new analysis, 
8 events are observed, while $7.8\pm 0.5$ background events are expected. 
For the favored LSND parameters, KARMEN would expect between 1.5 and 
5.5 oscillation events.  
The limit is now weaker than in Figure \ref{fig:karmen}.
These results encroach on 
LSND's allowed region, but they do not rule out LSND.
Another experiment will be needed to make a conclusive statement
about LSND, and that is where MiniBooNE comes in.

\section{MiniBooNE: experimental design}

MiniBooNE will use a new neutrino beamline coming
off of the FNAL 8 GeV proton Booster.  The Booster is a reliable
machine, expected to provide $2\times 10^7$ s of running per year, 
while delivering $5\times 10^{12}$ protons per 1 $\mu$s pulse
at a rate of 5 Hz to MiniBooNE.  Furthermore, the Booster will be able to 
deliver beam to MiniBooNE while also supplying protons for
the TeVatron and NuMI programs.

The secondary pion beam will emerge from a two-horn focusing 
system into a 50 m decay region.  The pion decay length will
be either 25 m or 50 m depending on the position of a movable
steel beam stop (varying the decay length provides a check of 
experimental systematics). The detector
will be positioned 500 m downstream of the decay region.

The detector
will consist of a spherical tank 6.1 m (20 feet) in radius
filled with 807 tons of pure mineral oil.
An inner-tank structure at 5.75 m will support phototubes and 
form an optical barrier, separating the tank into a central 
main volume and an outer veto shield. 
Cherenkov and scintillation light from neutrino interactions in 
the main volume will be detected by 1280 8-inch phototubes, providing
10\% photocathode coverage of the 445 ton fiducial volume.
(Undoped mineral oil tends to scintillate modestly
from the presence of intrinsic impurities.)
The veto shield 
will be viewed by 240 phototubes mounted on the tank wall.  

Typical neutrino energies will be from 0.5 to 1.0 GeV.
In one year of running, the experiment will collect approximately 
500000 reconstructed $\nu_\mu$ events. The intrinsic
$\nu_e$ contamination in the beam will be approximately 0.3\% 
or approximately 1500 reconstructed $\nu_e$ background events.

\section{MiniBooNE: analysis description}

The detector will reconstruct quasielastic $\nu_e$ 
interactions by identifying electrons via their characteristic
Cherenkov and scintillation light signatures. 
Besides the $\nu_e \rightarrow e^-$ signal, several backgrounds 
will contribute.  The analysis will come down to 
accounting for the backgrounds and determining whether or not 
there is an excess.
The background sources will be due to  $\nu_e$ contamination in the beam
and the misidentification of muons and $\pi^0$'s in the tank as
electrons.  
Because the neutrinos are at higher energies than at LSND and KARMEN, 
neutrons will not play a role in the signal and will not contribute
background.  

The detector will record the time of the initial hit and total charge 
for each phototube.   From this information, the track position
and direction will be determined.  Muon tracks will be distinguished 
from electron tracks by their Cherenkov rings and  
scintillation light.  Electrons will
tend to produce ``fuzzy'' rings due to multiple scattering and 
bremsstrahlung, while muon rings will tend to have sharp outer boundaries.
Electrons also tend to have a high fraction of prompt (Cherenkov)
light compared to late (scintillation) light, whereas muons 
produce relatively more late light.

The $\nu_e$ contamination in the beam is due to decays of pions 
and kaons.  Monte Carlo simulation
constrained by production data will be used to limit the systematic
uncertainty in the $\nu_e$ background to better than 10\%.  In addition, 
it will be possible to measure the pion energy spectrum using the 
the observed $\nu_\mu$ events, 
virtually all (99\%) of which will come from pion decay.
The technique 
exploits the classic energy-angle correlation in neutrino beams, 
which will be enhanced here by the relatively low beam energy and small
solid angle subtended by the detector.  By measuring the 
pion spectrum, MiniBooNE expects to
reduce the uncertainty in the pion component of the $\nu_e$ background
to less than 5\%.  

Ninety two percent of the muons contained in the detector will 
decay, and they will be relatively easily identified by the presence of 
a second track.  However, the 8\% of muons that get captured
have a greater chance to be misidentified.  
The misidentification of muon captures 
will be estimated by studying 
the large sample of muons that decay and determining the  
particle identification algorithm performance while ignoring the 
decay track.  Using this technique, which does not
rely on Monte Carlo simulation, the 
muon misidentification uncertainty is expected to be below 5\%.

Most neutral pions will be identified by their two electromagnetic 
decay tracks.  The small fraction (1\%) of asymmetric $\pi^0$ decays
will not yield two resolvable tracks and will therefore be more likely to
be misidentified. 
The misidentification contribution of these decays will be studied 
with Monte Carlo simulation, which will be constrained by 
the large sample of measured $\pi^0$'s in the experiment. 
The pion misidentification uncertainty is expected to be 5\%.  

\section{MiniBooNE: prospects}

\begin{figure}[b,t]
\centerline{\epsfxsize 3.0 truein \epsfbox{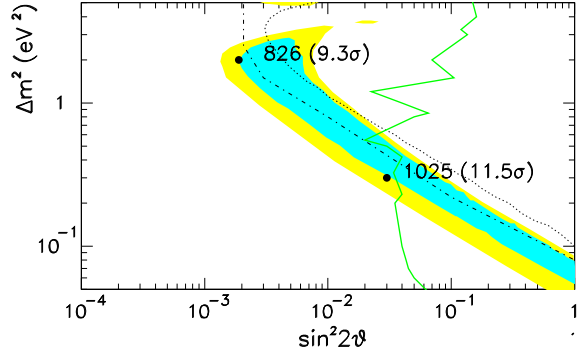}}
\caption[]{
\label{fig:boone1}
\small The expected number of excess events in MiniBooNE (and significance) 
for two points in the LSND favored region.}
\end{figure}

If oscillations occur as indicated by LSND, 
MiniBooNE will observe an excess of approximately 1000 events 
in one year of running.
Figure \ref{fig:boone1} 
shows the number of excess events and their significance for two points
in the LSND favored region. 
The significance is calculated using the systematic uncertainties 
for the various background sources above.

\begin{figure}[b,t]
\centerline{\epsfxsize 3.4 truein \epsfbox{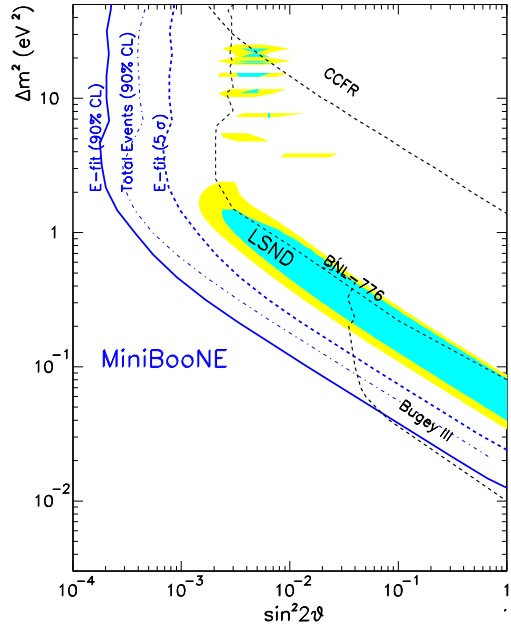}}
\caption[]{
\label{fig:boone}
\small MiniBooNE 90\% confidence level limits using energy-dependent 
fit (solid), and total event counting (dot-dash). 
Also shown is the $5\sigma$ sensitivity contour of the energy 
dependent fit (dashed).}
\end{figure}

MiniBooNE will gain additional sensitivity by measuring 
the energy dependence of the $\nu_e$ events.  The oscillation 
signal has a different energy distribution from the background. 
Therefore an underestimate of the background will not
necessarily lead to a fictitious oscillation signal.
Figure \ref{fig:boone} shows 
the exclusion contours 
for the energy-dependent fit as well as the limit  
based on using the total number of observed events above background.

\section{Conclusions}

LSND has presented evidence for muon to electron neutrino oscillations.
KARMEN is now searching for these oscillations, but KARMEN is unlikely
to have the sensitivity to reach a conclusion about LSND.  Another
experiment will be needed.

MiniBooNE is being prepared to fill this need. 
The detector and new 8 GeV beam line are being designed at Fermilab, 
and the experiment is scheduled to start data taking at the end of 2001.

MiniBooNE will either rule out LSND or it will demonstrate the signal
and home in on the parameters.  Should a single be found, BooNE would
be ready to continue its experimental program with a second detector, 
the position of which determined by the MiniBooNE result.

\section{Acknowledgements}

It is a pleasure to thank my MiniBooNE colleagues. I thank 
Eric Church for discussions 
about LSND and Klaus Eitel for discussions about KARMEN.


\begin{references}  

 \bibitem{church}Eric Church, these proceedings.

 \bibitem{lsnd1} C. Athanassopoulos, et al. (LSND), Phys. Rev. Lett. {\bf 75} (1995) 2650.

 \bibitem{karmen1} B. Zeitnitz et al. (KARMEN), {\it 19th Int. School of
 Nucl. Phys.} Erice, Italy, 1997, Prog. Part. Nucl. Phys. {\bf 40} (1998) 169.

 \bibitem{upgrade} G. Drexlin et al. (KARMEN), {\it 19th Int. School of
  Nucl. Phys.} Erice, Italy, 1997, Prog. Part. Nucl. Phys. {\bf 40} (1998) 193.

 \bibitem{karmen2} K. Eitel and B. Zeitnitz et al. (KARMEN), 
  {\it Neutrino'98}, Takayama, Japan, June 1998. 

 \bibitem{karmen3} T.E. Jannakos et al. (KARMEN), {\it Les 
  Rencontres de Moriond 1999}, Les Arcs, France, March 1999.
 
 \bibitem{yellin}S. J. Yellin, {\it COSMO-98}, 1999, hep-ex/9902012.  
 
 \bibitem{lsnddif} C. Athanassopoulos, et al. (LSND),  Phys. Rev. C 
 {\bf 58} (1998) 2489.
 

 \end{references}
\end{document}